\documentclass[10pt,twocolumn,superscriptaddress,prl]{revtex4-1}
\usepackage[latin9]{inputenc}
\usepackage{amsmath}
\usepackage{amssymb}
\usepackage{graphicx}

\makeatletter
\usepackage{amsfonts}
\usepackage{graphicx}
\usepackage{graphics}
\usepackage[usenames]{color}
\DeclareMathOperator{\Tr}{Tr}

\makeatother

\begin{document}

\title{Roton-induced Bose polaron in the presence of synthetic spin-orbit
coupling}

\author{Jia Wang}

\affiliation{Centre for Quantum and Optical Science, Swinburne University of Technology,
Melbourne 3122, Australia}

\author{Xia-Ji Liu}

\affiliation{Centre for Quantum and Optical Science, Swinburne University of Technology,
Melbourne 3122, Australia}

\author{Hui Hu}

\affiliation{Centre for Quantum and Optical Science, Swinburne University of Technology,
Melbourne 3122, Australia}

\date{\today}
\begin{abstract}
We theoretically investigate the quasiparticle (polaron) properties
of an impurity immersing in a Bose-Einstein condensate with equal
Rashba and Dresselhaus spin-orbit coupling at zero temperature. In
the presence of spin-orbit coupling, all bosons can condense into
a single plane-wave state with finite momentum, and the corresponding
excitation spectrum shows an intriguing roton minimum. We find that
the polaron properties are strongly modified by this roton minimum,
where the ground state of attractive polaron acquires a nonzero momentum
and anisotropic effective mass. Across the resonance of the interaction
between impurity and atoms, the polaron evolves into a tight-binding
dimer. We show that the evolution is not smooth when the roton structure
of the condensate becomes apparent, and a first-order phase transition
from a phonon-induced polaron to a roton-induced polaron is observed
at a critical interaction strength.
\end{abstract}
\maketitle
A central paradigm of quantum many-body theories is that elementary
excitations above a possibly strongly correlated ground state of a
quantum system can be approximated by well-defined quasiparticles.
In general, the nature and properties of quasiparticles can characterize
quantum phases, be predicted by many-body theories and be measured
by experiments \cite{WenBook}. Typically, a quantum phase transition
between ground states with different types of order manifests itself
as qualitative changes in the excitation spectrum, such as the appearance
of energy gaps \cite{SachdevBook}. The quasiparticle excitation spectrum
determined by spectral functions in many-body theories, therefore,
is a fundamental property of any interacting system \cite{Damascelli2003RMP,Stewart2008Nature}.
A mobile impurity interacting with a quantum-mechanical medium is
one of the most fundamental many-body systems in condensed matter
physics \cite{Landau1948JETP}. In such a system, the impurity is
dressed by coupling to the elementary excitations of the medium and
forms a polaron (that is a quasiparticle itself). The concept of polaron
has been systematically developed \cite{MahanBook} and provides deep
insights into quantum many-body physics. 

The great flexibility and extraordinary controllability make ultracold
atomic gases an excellent platform to study polaron physics \cite{Massignan2014RoPP}.
Fermi and Bose polarons, where impurity atoms are immersed respectively
in a degenerate Fermi gas \cite{Schirotzek2009PRL,Kohstall2012Nature,Koschorreck2012Nature,Zhang2012PRL,Scazza2017PRL}
and a Bose-Einstein condensate (BEC) \cite{HuMG2016PRL,Jorgensen2016PRL},
have both been experimentally realized. Ever since then, theoretical
studies have been proposed for a wide range of experimentally accessible
platforms: Fermi gases \cite{Chevy2006PRA,Lobo2006PRL,Combescot2007PRL,Punk2009PRA,Mathy2011PRL,Schmidt2012PRA},
Bose condensates \cite{Rath2013PRA,Shashi2014PRA,Li2014PRA}, Fermi
superfluids \cite{Nishida2015PRL,Yi2015PRA}, and long-range interacting
systems \cite{Kain2014PRA,WangJ2015PRL,SchmidtPRL2016,Camargo2018PRL}.
In particular, the tunability of interactions between impurities and
host atoms via Feshbach resonances has also stimulated a dramatic
theoretical improvement, allowing access to the strong-coupling regime
at resonance \cite{Levinsen2015PRL} and revealing physics beyond
perturbation treatments \cite{Astrakharchik2004PRA,Cucchietti2006PRL,Kalas2006PRA}.
It is now well understood that, even in the strong-coupling regime,
the properties of polaron can still be determined by modeling the
impurity coupling to only a few medium excitations: the particle-hole
excitations in the case of Fermi gases and the phonon excitations
in Bose condensates. By increasing the interaction further, Fermi
and Bose polarons turn into tightly bound dimers consisting of the
impurity and a single atom, via first- and second-order phase transitions,
respectively. In all cases, either polarons or dimers stay in a ground
state with zero center-of-mass momentum.

In this Letter, we report the existence of an exotic Bose polaron
induced by \emph{roton} excitations, which could emerge in superfluid
helium \cite{Greytak1969PRL}, quasi-two-dimensional dipolar BEC \cite{Chomaz2018NaturePhysics},
and spin-orbit coupled (SOC) BEC \cite{LiY2012PRL,Zhang2013JPB,Ji2015PRL},
all having a roton minimum in their excitation spectrum. Unlike the
conventional Bose polaron associated with phonon excitations, the
roton-induced Bose polaron features a ground state with \emph{finite}
center-of-mass momentum and \emph{anisotropic} effective mass, and
may also experience a \emph{first-order} phase transition towards
the dimer formation. We propose that this intriguing quasiparticle
could be readily observed in a weakly interacting $^{87}$Rb Bose
gas in the presence of synthetic SOC. 

To be specific, we focus here on the Raman-laser-induced SOC with
equal Rashba and Dresselhaus weight that has recently been successfully
engineered in ultracold quantum gases \cite{LinY2011Nautre,WangP2012PRL,Cheuk2012PRL,Dalibard2011PRL,Goldman2014RPP}.
At zero temperature $T=0$, a weakly interacting BEC with Raman SOC
can in principal exhibits three different quantum phases, namely the
stripe, plane-wave (PW) and zero-momentum (ZM) phases, depending on
the Rabi frequency $\Omega$ of the Raman lasers. While very interesting
properties, such as the emergence of density modulations in analogy
with supersolids \cite{WangC2010PRL,Ho2011PRL,LiY2012PRL,LiY2013PRL,LiY2014PRA,ChenXL2018PRA},
have been predicted in the stripe phase, its experimentally accessible
parameter space is narrow in typical BEC systems such as $^{87}$Rb.
This makes the observation of supersolid stripes very challenging
\cite{LiJ2016PRL,LiJ2017Nautre}. Here, we consider polarons in the
PW and ZM phases. It is in the PW phase that the Bogoliubov spectrum
has a local minimum at a finite quasi-momentum, which is attributed
as a roton minimum \cite{Zhang2013JPB}.

In greater detail, let us write down the Hamiltonian for the Raman
SOC BEC with atomic mass $m_{B}$, $\mathcal{H}_{{\rm BEC}}=\mathcal{H}_{0}+\mathcal{V}_{0}$,
where ($\hbar=1$ and the volume $V=1$) 
\begin{equation}
\mathcal{H}_{0}=\sum_{\mathbf{q}}\left[a_{\mathbf{q}\uparrow}^{\dagger},a_{\mathbf{q}\downarrow}^{\dagger}\right]\left[\frac{\left(\mathbf{q}-k_{{\rm SO}}\mathbf{\hat{e}}_{x}\sigma_{z}\right)^{2}}{2m_{B}}+\frac{\Omega}{2}\sigma_{x}\right]\left[\begin{array}{c}
a_{\mathbf{q\uparrow}}\\
a_{\mathbf{q}\downarrow}
\end{array}\right]
\end{equation}
and $\mathcal{V}_{0}=1/2\sum_{\mathbf{q},\mathbf{q',}\mathbf{k'};\sigma\sigma'}g_{\sigma\sigma'}a_{\mathbf{q'}\sigma}^{\dagger}a_{\mathbf{k'-q'}\sigma'}^{\dagger}a_{\mathbf{k'-q}\sigma'}a_{\mathbf{q}\sigma}$.
Here, $a_{\mathbf{q}\sigma}^{\dagger}$ ($a_{\mathbf{q}\sigma}$)
are creation (annihilation) operators of a boson with spin component
$\sigma=\{\uparrow,\downarrow\}$ and momentum $\mathbf{q}$, $k_{{\rm SO}}$
is the SOC strength defined by the recoil energy $E_{{\rm SO}}=k_{{\rm SO}}^{2}/(2m_{B})$,
and $\sigma_{x}$ and $\sigma_{z}$ are Pauli matrices. For $^{87}$Rb,
we take the SU(2)-invariant interaction $g_{\sigma\sigma'}=g_{B}\equiv4\pi a_{B}/m_{B}$,
$\text{\ensuremath{\forall\sigma},\ensuremath{\sigma}'}\in\{\uparrow,\downarrow\}$,
where $a_{B}$ is the BEC scattering length satisfying $0<a_{B}n^{-1/3}\ll1$
with $n$ being the BEC density.

\begin{figure}
\begin{centering}
\includegraphics[width=1\columnwidth]{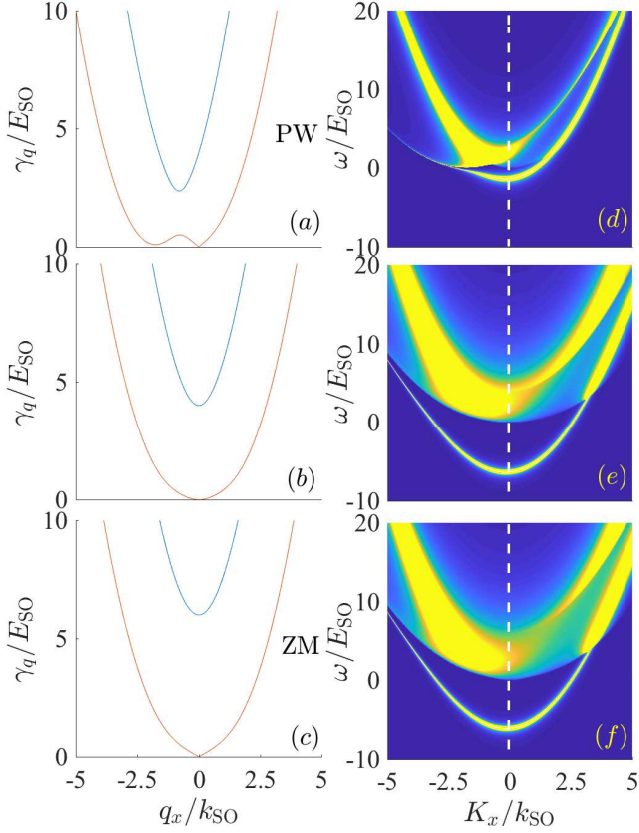}
\par\end{centering}
\caption{\textbf{Left panels:} Bogoliubov spectrum of a SOC BEC at $\Omega/E_{{\rm SO}}=2,\ 4,\ 6$,
with $g_{B}n=0.5E_{{\rm SO}}$ fixed. The two curves in each subplot
show the two branches of excitations. \textbf{Right panels:} The corresponding
spectral function for an impurity interacting with the BEC with $a_{\uparrow}=0$
and $a_{\downarrow}\rightarrow\infty$. The white dashed lines at
$K_{x}=0$ emphasize the asymmetry of the spectrum. Here and in all
the following plots, $\delta$-peaks are given a small artificial
width to be visible on the graph. \label{fig:BogoFig}}
\end{figure}

The mean-field ground state can be found via an ansatz that replaces
operators $a_{\mathbf{q}\sigma}^{\dagger}$ and $a_{\mathbf{q}\sigma}$
by complex numbers if and only if $\mathbf{q}=\mathbf{p}_{0}$, i.e.,
$a_{\mathbf{p}_{0}\uparrow}^{\dagger}=a_{\mathbf{p}_{0}\uparrow}=\phi_{\uparrow}\equiv\sqrt{n}\cos\theta$
and $a_{\mathbf{p}_{0}\downarrow}^{\dagger}=a_{\mathbf{p}_{0}\downarrow}=\phi_{\downarrow}\equiv-\sqrt{n}\sin\theta$
\cite{Zhang2013JPB}. Minimizing the total energy with respect to
variational parameters $\mathbf{p}_{0}$ and $\theta$ determines
the ground-state condensate wave function $\left[\phi_{\uparrow},\phi_{\downarrow}\right]$
and BEC chemical potential $\mu_{B}$ for a given $\Omega$. Two different
phases are separated by a critical Rabi frequency $\Omega_{c}=4E_{\textrm{SO}}$
\cite{Zhang2013JPB}: the PW phase for $\Omega<\Omega_{c}$ where
$\theta=0.5\sin^{-1}(\Omega/\Omega_{c})$ and $\mathbf{p}_{0}=k_{{\rm SO}}\cos2\theta\hat{\mathbf{e}}_{x}$;
and the ZM phase for $\Omega\ge\Omega_{c}$ where $\theta=\pi/4$
and $\mathbf{p}_{0}=\mathbf{0}$. These two phases have distinctive
quasiparticle spectra $\gamma_{\mathbf{q}}^{(b)}$ as shown in Fig.
\ref{fig:BogoFig}(a)-(c) for the chosen parameters $n=k_{{\rm SO}}^{3}$
and $g_{B}n=0.5E_{\textrm{SO}}$, all of which have upper and lower
branches denoted by the superscripts $b=\{+,-\}$, respectively. The
polaron properties at low energy are mainly determined by the lower
branch. As mentioned earlier, in the PW phase the lower branch of
Bogoliubov spectra exhibits a roton minimum near $\mathbf{p_{r}}=p_{r}\hat{\mathbf{e}}_{x}$,
as shown in Fig. \ref{fig:BogoFig}(a). The spectra are calculated
by applying Bogoliubov transformations $a_{\mathbf{\mathbf{p}_{0}+q,\sigma}}^{\dagger}=\sum_{b}[u_{\mathbf{q}\sigma}^{(b)}\beta_{\mathbf{q}}^{(b)\dagger}-v_{\mathbf{-q}\sigma}^{(b)}\beta_{-\mathbf{q}}^{(b)}]$
and $a_{\mathbf{p}_{0}-\mathbf{q,\sigma}}=\sum_{b}[u_{\mathbf{-q}\sigma}^{(b)}\beta_{\mathbf{-q}}^{(b)}-v_{\mathbf{q}\sigma}^{(b)}\beta_{\mathbf{q}}^{(b)\dagger}${]}
with $\beta_{\mathbf{q}}^{(b)\dagger}$ and $\beta_{\mathbf{q}}^{(b)}$
being the creation and annihilation operators of a Bogoliubov excitation.
Following Ref. \cite{Zhang2012PRL,ChenXL2017PRA}, we numerically
solve the Bogoliubov parameters $u_{\mathbf{q}\sigma}^{(b)}$, $v_{\mathbf{q}\sigma}^{(b)}$
and spectra $\gamma_{\mathbf{q}}^{(b)}$, to construct the 11-component
of the Green function, 
\begin{equation}
\left[G_{B}^{(11)}(\mathbf{q},i\nu_{n})\right]_{\sigma\sigma'}=\sum_{b}\left[\frac{u_{\mathbf{q}\sigma}^{(b)}u_{\mathbf{q}\sigma'}^{(b)}}{i\nu_{n}-\gamma_{\mathbf{q}}^{(b)}}-\frac{v_{\mathbf{-q}\sigma}^{(b)}v_{\mathbf{-q}\sigma'}^{(b)}}{i\nu_{n}+\gamma_{\mathbf{-q}}^{(b)}}\right],
\end{equation}
where $\nu_{n}=2n\pi k_{B}T$ are bosonic Matsubara frequencies.

By adding an impurity with mass $m_{I}$, the total model Hamiltonian
then becomes, $\mathcal{H}=\mathcal{H}_{{\rm BEC}}+\mathcal{H}_{I}+\mathcal{V}_{{\rm int}}$.
Here, the impurity Hamiltonian is simply $H_{I}=(\epsilon_{\mathbf{k}}-\mu_{I})c_{\mathbf{k}}^{\dagger}c_{\mathbf{k}}$,
where $\mu_{I}$ is the impurity chemical potential, $\epsilon_{\mathbf{k}}=k^{2}/(2m_{I})$
and $c_{\mathbf{k}}^{\dagger}\ (c_{\mathbf{k}})$ is the creation
(annihilation) operator of impurity. The interaction between impurity
and atoms is described by $\mathcal{V}_{{\rm int}}=\sum_{\sigma}g_{\sigma}^{(I)}\sum_{\mathbf{q},\mathbf{k},\mathbf{k'}}a_{\mathbf{k}\sigma}^{\dagger}c_{\mathbf{q}-\mathbf{k}}^{\dagger}c_{\mathbf{q}-\mathbf{k'}}a_{\mathbf{k'}\sigma},$
where the coupling constants $g_{\sigma}^{(I)}$ are regularized in
terms of the impurity-atom scattering lengths $a_{\sigma}$ and reduced
mass $\mu_{{\rm IB}}=m_{I}m_{B}/(m_{I}+m_{B})$: $[g_{\sigma}^{(I)}]^{-1}=\mu_{{\rm IB}}/2\pi a_{\sigma}-\sum_{\mathbf{k}}2\mu_{{\rm IB}}/k^{2}$,
so that strongly-interacting regime beyond perturbation regime can
be accessed. From now on, we focus on the case with $m_{I}=m_{B}=m_{0}$,
$a_{\uparrow}=0$ and $a_{\downarrow}\neq0$.

Without loss of generality, we assume that the impurity is fermionic,
and the bare impurity thermal Green function is given by $G_{I}^{(0)}(\mathbf{K},i\Omega_{m})=1/\left[i\Omega_{m}-(\epsilon_{\mathbf{K}}-\mu_{I})\right]$,
where $\Omega_{m}=(2m+1)\pi k_{B}T$ are fermionic Matsubara frequencies
and $\mathbf{K}$ denotes the center-of-mass momentum. We aim to calculate
the full Green's function incorporating the interaction with BEC 
\begin{equation}
G_{I}\left(\mathbf{K},i\Omega_{m}\right)=\frac{1}{i\Omega_{m}-\left(\epsilon_{\mathbf{K}}-\mu_{I}\right)-\Sigma_{I}\left(\mathbf{K},i\Omega_{m}\right)}\label{eq:ImpurityGF}
\end{equation}
that determines the spectral function $A(\mathbf{K},\omega)=-2\Im[G_{I}(\mathbf{K},i\Omega_{m}\rightarrow\omega+i0^{+}-\mu_{I})]$.
Here, for a weakly interacting BEC, $\Sigma_{I}(\mathbf{K},i\Omega_{m})\approx\Tr[n_{\mathbf{p}_{0}}\Gamma(\mathbf{K},i\Omega_{m})]$
is the self-energy, and $n_{\mathbf{p}_{0}}$ is a $2\times2$ matrix
with matrix elements $(n_{\mathbf{p}_{0}})_{\sigma\sigma'}=\phi_{\sigma}\phi_{\sigma'}$.
The vertex function is given by $\Gamma(\mathbf{K},i\Omega_{m})^{-1}=g_{I}^{-1}+\chi(\mathbf{K},i\Omega_{m})$,
where $g_{I}$ is a diagonal matrix with elements $g_{\sigma}^{(I)}$,
and the pair propagator $\chi(\mathbf{K},i\Omega_{m})=k_{B}T\sum_{\mathbf{\mathbf{q}},i\nu_{n}}G_{B}^{(11)}(\mathbf{q},i\nu_{n})G_{I}^{(0)}(\mathbf{K-\mathbf{q}},i\Omega_{m}-i\nu_{n})$.
The explicit expressions for the pair propagator $\chi(\mathbf{K},i\Omega_{m})$
and vertex function $\Gamma(\mathbf{K},i\Omega_{m})^{-1}$ as well
as the details of derivations are given in Supplemental Material \cite{SupplementalMaterial}. 

\begin{figure}
\begin{centering}
\includegraphics[width=1\columnwidth]{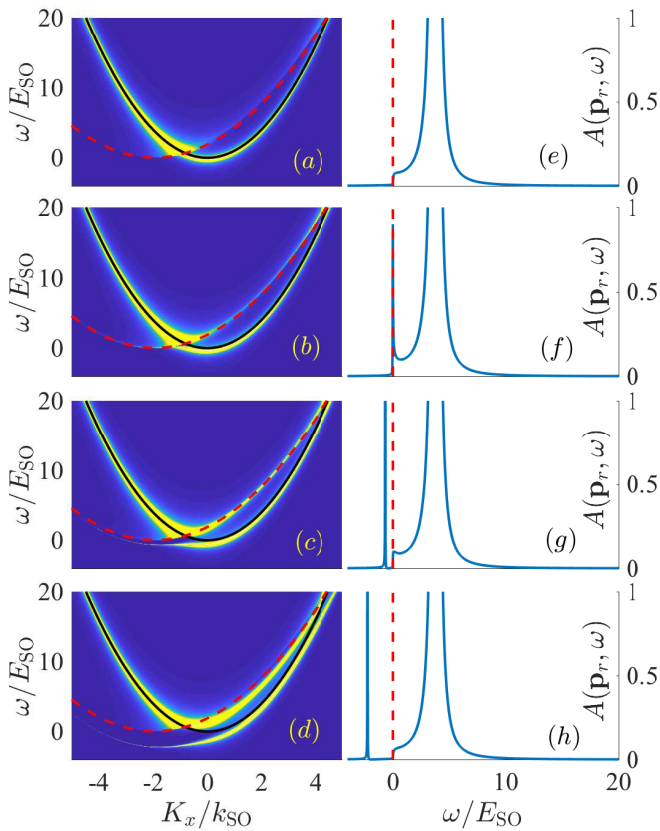}
\par\end{centering}
\caption{The spectral function for impurity atoms interacting with a SOC BEC
at the Rabi frequency $\Omega=E_{{\rm SO}}$. From top to bottom,
the parameters are chosen to be $1/(k_{{\rm SO}}a_{\downarrow})=-1,\ 0,\ 0.5,\ 1$,
with\textbf{ $a_{\uparrow}=0$ }and $g_{B}n=0.5E_{{\rm SO}}$ fixed.\textbf{
Left panels:} The frequency and momentum dependency of the spectral
function. The red dashed curve shows the dispersion of the center-of-mass
of a dimer consisting of a roton and an impurity, and the black solid curve
shows the dispersion of a free impurity atom $K_{x}^{2}/(2m_{0})$.
\textbf{Right panels:} The spectral function at roton minimum $\mathbf{p}_{r}=p_{r}\hat{\mathbf{e}}_{x}$
as a function of $\omega$. The red dashed line marks the roton minimum
energy $\Delta_{r}$. \label{fig:Avsa}}
\end{figure}

The frequency and ($K_{x}$, the $x$-component) momentum dependence
of the spectral functions are shown in Fig. \ref{fig:BogoFig}(d)-(f)
for $\Omega=2E_{{\rm SO}}$ (PW), $4E_{{\rm SO}}$ (critical) and
$6E_{{\rm SO}}$ (ZM) respectively. The spectral function splits into
two branches (similar to the polaron spectrum in a conventional BEC).
The upper (lower) branch is called a repulsive (attractive) polaron.
The repulsive polaron corresponds to a resonance with broad width
at high energy. In contrast, the attractive polaron is a well-defined
quasiparticle at small momentum and the corresponding spectral function
is proportional to a delta function (On the graph, for visibility
this $\delta$-peak is given a small artificial width). One can observe
that, the polaron spectrum is not symmetric (with respect to $K_{x}\rightarrow-K_{x}$)
even in the ZM phase where $\gamma_{\mathbf{q}}^{(b)}$ are symmetric.
This can be understood by realizing that $u_{\mathbf{q}}^{(b)}$ and
$v_{\mathbf{q}}^{(b)}$ are not symmetric, and the impurity is only
interacting with the spin-down component for our choosing parameters
($a_{\uparrow}=0$). This asymmetry of polaron spectrum leads to a
minimum of the attractive polaron energy at a non-zero momentum. In
another word, the ground state of Bose polaron in the presence of
SOC acquires a finite momentum $\mathbf{K}^{*}=K_{x}^{*}\hat{\mathbf{e}}_{x}$.
At resonance $a_{\downarrow}\rightarrow\infty$, $\mathbf{K^{*}}$
is much smaller than the roton minimum momentum $\mathbf{p_{r}}$,
implying the formation of the polaron ground state is mostly contributed
by phonons (the linear dispersion of Bogoliubov spectrum near origin)
instead of rotons (the local minimum near $\mathbf{p_{r}}$). Nevertheless,
Fig. \ref{fig:BogoFig}(a) shows that the spectral function has a
rich and interesting structure near $\mathbf{p_{r}}$, indicating
that the roton minimum in the PW phase strongly modifies the excitation
spectrum of polaron at resonance. 

To understand the role of rotons better, it is instructive to see
how the quasiparticle properties of the polaron changes as the impurity-BEC
interaction $a_{\downarrow}$ is varied. As $a_{\downarrow}$ varies
across the resonance, an attractive Bose polaron evolves to a tight-binding
dimer. Intuitively, if the roton minimum is close enough to zero,
we expect that the dimer could be strongly affected by the roton minimum
and the polaron spectrum might also be strongly modified. This evolution
is shown in Fig. \ref{fig:Avsa} for polarons in the deep PW phase
at $\Omega=E_{{\rm SO}}$. For this Rabi frequency, the lower branch
of BEC Bogoliubov dispersion $\gamma_{q}^{(-)}$ near the roton minimum
along $\hat{x}$ can be approximated by $\gamma_{q}^{(-)}\approx(q_{x}-p_{r})^{2}/m_{x}^{(r)}+\Delta_{r}$,
where $p_{r}\approx-1.947k_{{\rm SO}}$, $\Delta_{r}\approx0.025E_{{\rm SO}}$
and $m_{x}^{(r)}=1.041m_{0}$ can be obtained by numerical fitting.
In Fig. \ref{fig:Avsa}, we set $(k_{{\rm SO}}a_{\downarrow})^{-1}=-1,\ 0,\ 0.5$
and $1$ from top to bottom. The left panels show the frequency and
momentum dependency of spectral functions. The polaron can be understood
as a modification of free impurity by dressing medium excitations
at perturbation region, and hence the spectral function shows only
a single branch of resonances near free-particle dispersions indicated
by the black solid curve in Fig. \ref{fig:Avsa}(a). We also find
that this spectral function acquires a sizable width in regimes above
a red dashed threshold curve corresponding to the dispersion of the
center-of-mass of a dimer consisting of a roton and an impurity, which
along the $x$-axis is given by $(K_{x}-p_{r})^{2}/2[m_{I}+m_{x}^{(r)}]+\Delta_{r}$.
Above this threshold, the polaron scatters off virtual roton-impurity
dimers and suffers a finite lifetime. At the unitarity $a_{\downarrow}\rightarrow\infty$,
sharp peaks show up near the roton-impurity dimer center-of-mass dispersions
in Fig. \ref{fig:Avsa}(b), and forms an ``avoid-cross'' at positive
intermediate interaction $a_{\downarrow}^{-1}=0.5k_{{\rm SO}}$ in
Fig. \ref{fig:Avsa}(c). At $a_{\downarrow}^{-1}=k_{{\rm SO}}$ in
Fig. \ref{fig:Avsa}(d), the polaron spectrum become two well separated
branches. These behaviors are emphasized in Fig. \ref{fig:Avsa}(e)-(h)
for the spectral function at roton minimum momentum $K_{x}=q_{r}\hat{\mathbf{e}}_{x}$.
Another peculiar feature for the attractive polaron (lower branches)
is the formation of a double-well structure in Fig. \ref{fig:Avsa}(c),
suggesting that a first-order phase transition is possible in the
deep PW phase regime when the Rabi frequency is smaller than a threshold
$\Omega^{*}$. Our numerical indicates that $\Omega^{*}=1.2E_{{\rm SO}}$
for our choosing parameters.

\begin{figure}
\begin{centering}
\includegraphics[width=1\columnwidth]{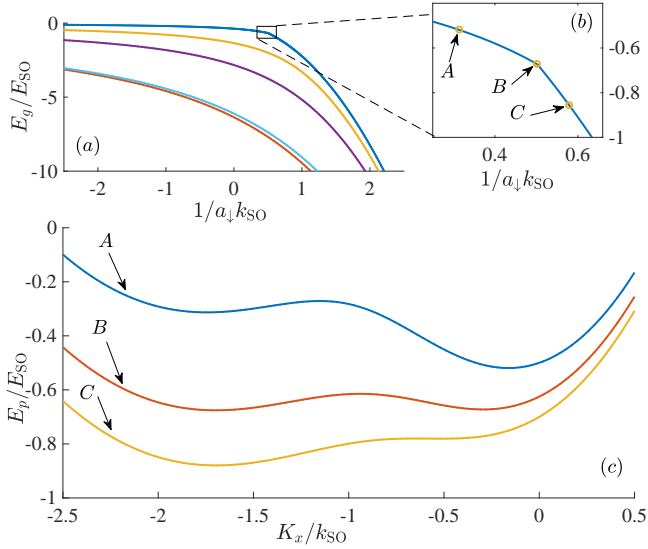}
\par\end{centering}
\caption{(a) Attractive polaron ground state energy $E_{g}$ as a function
of $1/a_{\downarrow}$ at different Rabi frequency $\Omega/E_{{\rm SO}}=1,2,3,4$
and $6$ from top to bottom. We fixed the parameter $a_{\uparrow}=0$
and $g_{B}n=0.5E_{{\rm SO}}$ here. (b) A zoom-in for polaron energy
$\Omega=E_{{\rm SO}}$ to emphasize the phase transition. (c) The
double minimum structure of attractive polaron energy $E_{p}(K_{x}\hat{\mathbf{e}}_{x})$
as a function of $K_{x}$ at point $A,B$ and $C$ in (b). \label{fig:AttrPolaron}}
\end{figure}

The double minimum for attractive polaron at $\Omega=E_{{\rm SO}}$
can be studied qualitatively by examining the polaron energy spectrum
given as the pole of the Green function Eq. (\ref{eq:ImpurityGF}),
i.e., $E_{p}(\mathbf{K})\approx\epsilon_{\mathbf{K}}+\Tr[n_{\mathbf{p}_{0}}\Gamma(\mathbf{K},E_{p}(\mathbf{K})-\mu_{I}+i0^{+})]$.
We define the attractive polaron ground-state energy $E_{g}$ as the
global minimum of $E_{p}(\mathbf{K})$ at momentum $\mathbf{K^{*}}=K_{x}^{*}\hat{\mathbf{e}}_{x}$.
In Fig. \ref{fig:AttrPolaron}(c), one can clearly see the global
minimum change from the one near the origin (that we called phonon-induced
polaron) to the one near $p_{r}$ (roton-induced polaron). The polaron
energy $E_{g}$ is not smooth as a function of $1/a_{\downarrow}$
at the point $B$ in Fig. \ref{fig:AttrPolaron}(b), indicating a
first-order phase transition between the phonon-induced and roton-induced
polaron. This is in comparison with polaron energy for higher $\Omega>\Omega^{*}$
in Fig. \ref{fig:AttrPolaron}(a), which are continuous and smooth.

\begin{figure}
\begin{centering}
\includegraphics[width=1\columnwidth]{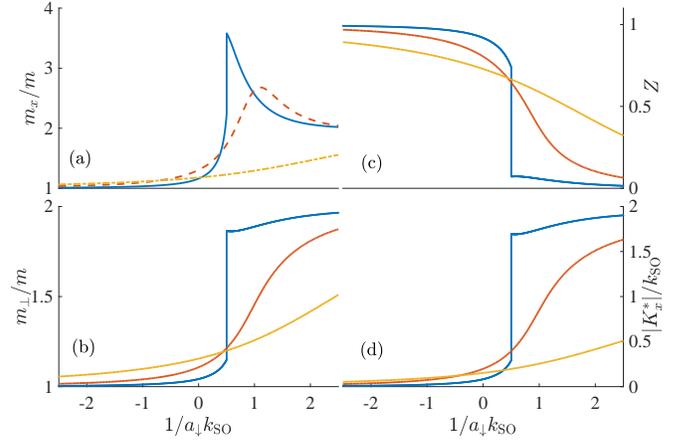}
\par\end{centering}
\caption{(a) Effective mass $m_{x}$, (b) $m_{\perp}$ (c) spectral residue
$Z$ and (d) the center-of-mass momentum $K_{x}^{*}$ of attractive
polaron, as a function of $1/a_{\downarrow}$ at different Rabi frequency
$\Omega/E_{{\rm SO}}=1,2$ and 6, indicated by blue solid, red dashed
and yellow dash-dotted curves correspondingly. \label{fig:APolaronProperty}}
\end{figure}

We now address other polaron properties near the global minimum $\mathbf{K^{*}}$
by approximating the Green function near the pole $E_{p}(\mathbf{K}^{*})$
to \cite{Rath2013PRA}
\begin{equation}
G_{I}(\mathbf{K},\omega-\mu_{I})\approx\frac{Z}{\omega-E_{p}(\mathbf{K}^{*})-\frac{\hbar^{2}\left(K_{x}-K_{x}^{*}\right)^{2}}{2m_{x}}-\frac{\hbar^{2}K_{\perp}^{2}}{2m_{\perp}}},
\end{equation}
where $Z^{-1}=1-\partial_{\Omega}\left.\sum_{I}(\mathbf{K},\Omega)\right|_{\mathbf{K\rightarrow\mathbf{K^{*}}},\Omega\rightarrow E_{p}(\mathbf{K^{*}})-\mu_{I}}$
is the spectral residue. The effective mass in $x$-direction $m_{x}$
and the perpendicular direction $m_{\perp}$ are given by $Z^{-1}m_{i}^{-1}=m_{0}^{-1}+\left.\partial^{2}\Sigma_{I}\left(\mathbf{K},\Omega\right)/\partial K_{i}^{2}\right|_{\mathbf{K\rightarrow K^{*},}\Omega\rightarrow E_{p}(\mathbf{K^{*}})-\mu_{I}}$,
and are in general anisotropic, $m_{x}\neq m_{\perp}$, as shown in
Fig. \ref{fig:APolaronProperty}(a) and (b) for $\Omega/E_{{\rm SO}}=1,2$
and 6. We also present the spectral residue $Z$ and center-of-mass
momentum $K_{x}^{*}$ in Fig. \ref{fig:APolaronProperty}(c) and (d).
All these quantities show discontinuity for the case $\Omega=E_{{\rm SO}}$,
as a result of the first-order transition between roton-induced polaron
and dimer. For $\Omega=2E_{{\rm SO}}$ and $6E_{{\rm SO}}$, instead,
they change rather smoothly, following the conventional second-order
transition from a phonon-induced polaron to a dimer \cite{Rath2013PRA}.
The first-order transition found here could be related to the parabolic
dispersion of the roton. A similar first-order transition is also
observed for Fermi polarons, where the dispersion of medium excitations
(i.e., particle-hole excitations) is parabolic. In contrast, the phonon
dispersion is linear, leading to a smooth polaron-dimer transition
for phonon-induced Bose polarons.

In Fig. \ref{fig:APolaronProperty}(a), the effective mass $m_{x}$
shows a bump near $1/a_{\downarrow}\approx k_{{\rm SO}}$ at $\Omega=2E_{{\rm SO}}$.
This strong deviation of $m_{x}$ from $2m$ shows that the polaron
is a genuine many-body effect. Nevertheless, our numerical calculation
indicates that this bump become gradually less significant and disappear
for larger $\Omega$ but before $\Omega_{c}$ (not shown here), suggesting
that the existence of the bump is not a good indicator of the PW phase
of the host medium BEC.

The polaron phenomena studied here is experimentally accessible. We
may consider a $^{87}$Rb condensate with scattering length $a_{B}=100a_{0}$.
For a laser-induced SOC, the typical SOC strength is set to be $k_{{\rm SO}}=2\pi\times10^{6}$
m$^{-1}$. Our chosen parameter $gn=0.5E_{{\rm SO}}$ then corresponds
to a BEC density $n\simeq3\times10^{14}$ cm$^{-3}$. The impurity
can be a $^{87}$Rb atom in another hyperfine state that is not coupled
by the SOC lasers, and the interaction between the impurity and the
SOC BEC can be easily tuned by using Feshbach resonances.

In summary, we have predicted the emergence of an exotic quasiparticle
- roton-induced Bose polaron - in a spin-orbit coupled Bose-Einstein
condensate, which has a nonzero center-of-mass momentum and anisotropic
effective masses, and undergoes a first-order polaron-molecule transition
upon varying impurity-atom interaction. These unusual properties
may also be examined in dipolar Bose condensate and superfluid helium,
when rotons are thermally excited at finite temperature.

We are grateful to Professor Zeng-Qiang Yu and Brendan C. Mulkerin
for useful discussions. This research was supported by the Australian
Research Council (ARC) Discovery Programs, Grants No. DE180100592
and No. DP190100815 (J.W.), No. FT140100003 and No. DP180102018 (X.-J.L),
and No. DP170104008 (H.H.).

\bibliographystyle{apsrev4-1}
\bibliography{RefsRotonPolaron}
\pagebreak
\begin{widetext}
\section{Many-body $T$-matrix theory of Bose polarons}

Here we generalize the many-body $T$-matrix theory of Bose polarons
in Ref. \cite{Rath2013PRA} to a spin-orbit coupled Bose-Einstein
condensate (BEC). We have also performed numerical calculations by
using Chevy's variational approach \cite{Chevy2006PRA,Li2014PRA},
which yields essentially the same results.

\subsection{BEC Green function}

We construct the zero-temperature Green function of the BEC atoms
by using the Bogoliubov parameters $u_{\mathbf{q}\sigma}^{(b)}$,
$v_{\mathbf{q}\sigma}^{(b)}$ and Bogoliubov spectra $\gamma_{\mathbf{q}}^{(b)}$,
\begin{equation}
\mathbf{G}_{B}(\mathbf{q},i\nu_{n})=\left[\begin{array}{cc}
G_{B}^{(11)}(\mathbf{q},i\nu_{n}) & G_{B}^{(12)}(\mathbf{q},i\nu_{n})\\
G_{B}^{(21)}(\mathbf{q},i\nu_{n}) & G_{B}^{(22)}(\mathbf{q},i\nu_{n})
\end{array}\right],
\end{equation}
where $G_{B}^{(ij)}$ are two-by-two matrices, whose matrix elements
can be explicitly written out as ($\sigma,\sigma'=\uparrow,\downarrow$)
\cite{Zhang2013JPB},

\begin{equation}
\left[G_{B}^{(11)}(\mathbf{q},i\nu_{n})\right]_{\sigma\sigma'}=\left[G_{B}^{(22)}(\mathbf{-q},-i\nu_{n})\right]_{\sigma\sigma'}=\sum_{b}\left[\frac{u_{\mathbf{q}\sigma}^{(b)}u_{\mathbf{q}\sigma'}^{(b)}}{i\nu_{n}-\gamma_{\mathbf{q}}^{(b)}}-\frac{v_{\mathbf{-q}\sigma}^{(b)}v_{\mathbf{-q}\sigma'}^{(b)}}{i\nu_{n}+\gamma_{\mathbf{-q}}^{(b)}}\right],
\end{equation}

\begin{equation}
\left[G_{B}^{(12)}(\mathbf{q},i\nu_{n})\right]_{\sigma\sigma'}=\left[G_{B}^{(21)}(-\mathbf{q},-i\nu_{n})\right]_{\sigma\sigma'}=\sum_{b}\left[-\frac{u_{\mathbf{q}\sigma}^{(b)}v_{\mathbf{q}\sigma'}^{(b)}}{i\nu_{n}-\gamma_{\mathbf{q}}^{(b)}}+\frac{v_{\mathbf{-q}\sigma}^{(b)}u_{\mathbf{-q}\sigma'}^{(b)}}{i\nu_{n}+\gamma_{\mathbf{-q}}^{(b)}}\right].
\end{equation}
where $\nu_{n}=2n\pi k_{B}T$ are bosonic Matsubara frequencies. As
shown below, in our diagrammatic treatment within ladder approximation,
only the 11-component of the propagator $G_{B}^{(11)}$ is needed
to calculate the single-particle spectral function of Bose polarons
at zero temperature.

\subsection{Many-body $T$-matrix theory}

Without loss of generality, we assume the impurity is fermionic, with
a bare thermal Green function given by $G_{I}^{(0)}(\mathbf{K},i\Omega_{m})=1/\left[i\Omega_{m}-(\epsilon_{\mathbf{K}}-\mu_{I})\right]$,
where $\Omega_{m}=(2m+1)\pi k_{B}T$ are fermionic Matsubara frequencies
and $\mathbf{K}$ denotes the center-of-mass momentum. We aim to calculate
the thermal Green function including the interaction with BEC, 
\begin{equation}
G_{I}(\mathbf{K},i\Omega_{m})=\frac{1}{i\Omega_{m}-(\epsilon_{\mathbf{K}}-\mu_{I})-\Sigma_{I}(\mathbf{K},i\Omega_{m})},
\end{equation}
where the self-energy $\Sigma_{I}(\mathbf{K},i\Omega_{m})$ calculated
within ladder approximation is given by \cite{Rath2013PRA} 
\begin{equation}
\Sigma_{I}(\mathbf{K},i\Omega_{m})=\Tr[n_{\mathbf{p}_{0}}\Gamma(\mathbf{K},i\Omega_{m})]+\Sigma_{I}^{\mathrm{ex}}(\mathbf{K},i\Omega_{m}).
\end{equation}
Here $n_{\mathbf{p}_{0}}$ is a $2\times2$ matrix with elements $(n_{\mathbf{p}_{0}})_{\sigma\sigma'}=\phi_{\sigma}\phi_{\sigma'}$,
$\Gamma(\mathbf{K},i\Omega_{m})$ is the vertex function, and 
\begin{equation}
\Sigma_{I}^{\mathrm{ex}}(\mathbf{K},i\Omega_{m})=-k_{B}T\sum_{\mathbf{q},i\nu_{n}}\Tr\left[G_{B}^{(11)}(\mathbf{q},i\nu_{n})\Gamma(\mathbf{K}+\mathbf{q},i\nu_{n}+i\Omega_{m})\right].
\end{equation}
At zero temperature, the Matsubara summation can be carried out by
neglecting contributions from the pole of vertex function $\Gamma$.
This pole corresponds to the energy of a dimer or molecule state,
whose macroscopic occupation is vanishingly small at zero temperature.
The Matsubara summation therefore gives
\begin{equation}
\Sigma_{I}^{\mathrm{ex}}(\mathbf{K},i\Omega_{m})=\sum_{\mathbf{q},b}\Tr\left[n_{\mathrm{\mathrm{ex}}}^{(b)}(\mathbf{q})\Gamma(\mathbf{K+\mathbf{q}},i\Omega_{m}-\gamma_{\mathbf{k}}^{(b)})\right],
\end{equation}
where $n_{\mathrm{\mathrm{ex}}}^{(b)}(\mathbf{q})$ is a $2\times2$
matrix with matrix element: $[n_{\mathrm{\mathrm{ex}}}^{(b)}(\mathbf{q})]_{\sigma\sigma'}=v_{-\mathbf{q}\sigma}^{(b)}v_{-\mathbf{q}\sigma'}^{(b)}$,
implying that this part of self-energy is determined by \emph{quantum
depletion} and is negligible when the boson-boson interactions in
the host medium are weak. Therefore, to a good approximation we obtain,
\begin{equation}
\Sigma_{I}(\mathbf{K},i\Omega_{m})=\Tr[n_{\mathbf{p}_{0}}\Gamma(\mathbf{K},i\Omega_{m})].
\end{equation}

\subsection{The vertex function}

The vertex function is given by 
\begin{equation}
\Gamma(\mathbf{K},i\Omega_{m})^{-1}=g_{I}^{-1}+\chi(\mathbf{K},i\Omega_{m}),
\end{equation}
where $g_{I}$ is a diagonal matrix with elements $g_{\sigma}^{(I)}$,
and the pair propagator takes the form, 
\begin{equation}
\chi(\mathbf{K},i\Omega_{m})=k_{B}T\sum_{\mathbf{\mathbf{q}},i\nu_{n}}G_{B}^{(11)}(\mathbf{q},i\nu_{n})G_{I}^{(0)}(\mathbf{K-\mathbf{q}},i\Omega_{m}-i\nu_{n}).
\end{equation}
The Matsubara summation can be carried out analytically and gives
explicit expressions of matrix elements at zero temperature limit,
\begin{equation}
\left[\chi(\mathbf{K},i\Omega_{m})\right]_{\sigma\sigma'}=-\sum_{\mathbf{q},b}\frac{u_{\mathbf{\mathbf{q}\sigma}}^{(b)}u_{\mathbf{q\sigma'}}^{(b)}}{i\Omega_{m}-\gamma_{\mathbf{q}}-(\epsilon_{\mathbf{K-\mathbf{q}}}-\mu_{I})}.
\end{equation}
It can be shown that the diagonal matrix elements have an ultraviolet
divergence that can be removed by the renormalization of coupling
constants $g_{\sigma}^{(I)}$, i.e. ($m_{B}=m_{I}=m$),
\begin{equation}
\left[\Gamma^{-1}(\mathbf{K},i\Omega_{m})\right]_{\sigma\sigma'}=\frac{m}{4\pi a_{\sigma}}\delta_{\sigma\sigma'}-\sum_{\mathbf{q},b}\left[\frac{u_{\mathbf{q\sigma}}^{(b)}u_{\mathbf{q\sigma'}}^{(b)}}{i\Omega_{m}-\gamma_{\mathbf{q}}-(\epsilon_{\mathbf{K-\mathbf{q}}}-\mu_{I})}+\frac{m}{q^{2}}\delta_{\sigma\sigma'}\right].
\end{equation}

By taking analytic continuation $i\Omega_{m}\rightarrow\omega-\mu_{I}+i0^{+}$,
we have the expression for the polaron energy $E_{p}(\mathbf{K})$
as the pole of the Green function $G_{I}$,
\begin{eqnarray}
E_{p}\left(\mathbf{K}\right) & \simeq & \epsilon_{\mathbf{K}}+\Tr\left[n_{\mathbf{p}_{0}}\Gamma\left(\mathbf{K},E_{p}-\mu_{I}+i0^{+}\right)\right],
\end{eqnarray}
 which is ready to calculated numerically.

\section{Evolution of the polaron spectral function across the impurity-atom
Feshbach resonance}

To visualize the evolution of the polaron spectral function $A(K_{x}\hat{\mathbf{e}}_{x},\omega)$
as a function of $1/(k_{\textrm{SO}}a_{\downarrow})$ at $a_{\uparrow}=0$,
we make three animations for the plane-wave phase ($\Omega=E_{\textrm{SO}}$),
critical point ($\Omega=4E_{\textrm{SO}}$), and zero-momentum phase
($\Omega=6E_{\textrm{SO}}$). They are recorded in the files, SOCBECPolaron1.mov,
SOCBECPolaron4.mov and SOCBECPolaron6.mov, respectively and can be provided by request.

\end{widetext}

\end{document}